\documentclass[prb,aps,twocolumn,showpacs]{revtex4}

\usepackage{epsfig}
\usepackage{bm}

\def\vecb#1{{\bf #1}}

\begin{document}

\title{Quantum phase transition from an antiferromagnet to a spin liquid in a metal}

\author{Tarun Grover}
\affiliation{Department of Physics, Massachusetts Institute of
Technology, Cambridge, Massachusetts 02139}
\author{T. Senthil}
\affiliation{Department of Physics, Massachusetts Institute of
Technology, Cambridge, Massachusetts 02139}

\begin{abstract}
We study quantum phase transitions from easy-plane antiferromagnetic metals to paramagnetic metals in
Kondo-Heisenberg lattice systems. If the paramagnetic metal is a fractionalized Fermi liquid
then the universal critical properties of the phase transition are unaffected for a weak Kondo coupling even when the Fermi surface intersects the magnetic zone boundary. 
 This is in striking contrast to the conventional theory of phase transitions between paramagnetic and antiferromagnetic metals where any Kondo coupling is strongly relevant, and leads to a Landau-damped `Hertz-Millis' theory. The electron quasi-particle remains well-defined in the quantum critical regime and the critical spin fluctuations only contribute subleading corrections to the various properties of conduction electrons.

\end{abstract}


\maketitle

\newcommand{\be}{\begin{equation}}
\newcommand{\ee}{\end{equation}}
\newcommand{\bea}{\begin{eqnarray}}
\newcommand{\eea}{\end{eqnarray}}
\newcommand{\bK}{\textbf{K}}
\newcommand{\bp}{\textbf{p}}
\newcommand{\kw}{\bm{k},\omega}

\section{Introduction}

The earliest and most influential approach to studying  magnetic quantum phase transitions in metals (developed by Hertz, Moriya, Millis \cite{hertz76} and others (for additional references see e.g. \cite{lohrmp}) focuses on the
fluctuations of the natural magnetic order parameter. The metallic environment results in damping of these magnetic fluctuations. The resulting theory is however
weakly coupled in physical dimension $d \geq 2$ - consequently it allows only for weak deviations from Fermi liquid physics even right at the magnetic quantum critical point. In particular the spin fluctuation spectrum at finite temperature $T$ and a frequency $\omega$ does not show scaling as a function of $\omega/T$.

Experimentally the best studied magnetic quantum phase transitions are in heavy fermion metals \cite{lohrmp} which have a lattice of local moments coupled
via Kondo exchange to a partially filled band of conduction electrons. Remarkably in many cases striking breakdown of Fermi liquid physics is observed near these quantum critical points which is not understood even qualitatively. For instance the spin fluctuation spectrum in $CeCu_{6-x}Au_x$ at quantum criticality
is scale invariant and shows $\omega/T$ scaling \cite{stock98, schr98, schr00}. A number of transport and thermodynamic anomalies have also been seen in this and other materials \cite{loh94, loh96, loh98}. In the heavy fermion context
the Hertz-Moriya-Millis (HMM) approach treats the magnetism as a spin density wave instability of the heavy Fermi liquid and essentially assumes that the Kondo process responsible for the formation of the Fermi liquid state continues across the magnetic transition.
An alternate approach specific to heavy electron metals suggests that the transition to magnetism is accompanied by the breakdown of the Kondo effect \cite{si0103,coleman01,senthil03,senthil04}. Empirically this will manifest itself as a reconstruction of the electronic Fermi surface associated with the loss of participation of the local moments in the Fermi sea. Evidence in favor of such a Fermi surface reconstruction accompanying the magnetic ordering transition has been presented in $YbRh_2Si_2$ \cite{pasch04} and in $CeRhIn_5$\cite{shish05}. Very recently studies of $YbRh_2Si_2$ with $Co$ or $Ir$ substituted at the $Rh$ site have suggested that the Fermi surface reconstruction may be separated from the magnetic transition through chemical pressure \cite{fried09}. Most interestingly with $Ir$ substitution, the antiferromagnetic metal appears to be separated from the paramagnetic large Fermi surface metal through an intermediate phase which has neither magnetism nor Kondo screening. The nature of this intermediate phase and the associated quantum phase transitions remains to be elucidated.

On the theoretical side much attention has focused on the possible quantum phases and phase transitions of Kondo lattices driven by competition between inter-moment exchange and the Kondo effect. Two common phases are of course a magnetically ordered metal with a ``small" conduction electron Fermi surface, and a paramagnetic heavy Fermi liquid with a ``large" Fermi surface. An intriguing third possibility is that the local moments form a quantum spin liquid state which then remains decoupled from the conduction electrons\cite{burdin02,senthil03,senthil04}. The resulting paramagnetic metal will have a small Fermi surface and hence violate the usual Luttinger theorem\cite{senthil03}. Such phases were dubbed `fractionalized Fermi liquids' and denoted FL$^*$. Quantum phase transitions in Kondo lattice systems are much less understood. The possibility of a single direct transition where the magnetic ordering is accompanied by the destruction of the Kondo effect (and the associated Fermi surface reconstruction) has been discussed but without satisfactory resolution. Progress has been possible in analyzing phase transitions associated with the breakdown of the Kondo effect without worrying about the magnetic ordering. A concrete theory of such a `Kondo breakdown' transition from the usual heavy Fermi liquid to a `small Fermi surface' metal was proposed in Ref. \cite{svs} and further developed in Refs. \cite{kb}.

In this paper we explore instead the complementary problem of the quantum phase transition associated with the destruction of magnetism without worrying about the onset of Kondo screening. Specifically we study the phase transition between a magnetic metal and a particular fractionalized Fermi liquid where the local moments form a quantum spin liquid. We focus on the interesting case where the magnetic ordering wave vector can connect distinct points of the conduction electron Fermi surface. Remarkably we find that the magnetic phase transition remains strongly coupled, and satisfies nontrivial scaling (such as in the spin fluctuation spectrum as a function of $\omega/T$). This is in striking contrast to the usual HMM spin density wave transition. We also study the effects of these critical spin fluctuations on the properties of the electronic quasiparticles of the quantum critical metal.

Our considerations will be based on Kondo-Heisenberg models with the general structure
\begin{eqnarray}
H & = & H_c + H_K + H_{int} \\
H_c & = & \sum_k \epsilon_k c^{\dagger}_k c_k \\
H_K & = & \frac{J_K}{2} \sum_r \vec S_r. c^{\dagger}_r \vec \sigma c_r \\ \label{kndo}
H_{int} & = & \sum_{rr'} J_a S^a_r S^a_{r'} + ....
\end{eqnarray}
Here $c_{k\alpha}$ destroys a conduction electron of momentum $k$ and spin $\alpha$ and $\vec S_r$ are spin-$1/2$ operators describing the
localized moments.  In the last term we will allow for
other short ranged interactions between the local moments that enable tuning the local moment system through magnetic phase transitions.

Generally the magnetic phase transitions of the kind we wish to study are conveniently accessed theoretically by first ignoring any coupling between conduction electrons and the local moments. Then the local moment system is insulating and much is known about the phases and phase transitions of such quantum magnets. The general question we are concerned with is the effect of a small coupling of this insulating quantum magnet to the metallic electrons. The best understood example of such a quantum phase transition in an insulating magnet is that between a Neel state and a gapped quantum spin liquid.
In the simplest such spin liquid state the excitations consist of
spin-$1/2$ bosonic spinons. In addition there are gapped $Z_2$ vortices (visons)  that act as sources of $\pi$ flux for the spinons.
Second order quantum phase transitions to such states from magnetically ordered states are known to be possible in two cases - first for magnets with noncollinear
magnetization patterns and second for magnets with easy plane anisotropy with either collinear or noncollinear magnetic order.  For concreteness we will focus on easy plane quantum antiferromagnets with collinear order in this paper.

\begin{figure}[tb]
\centerline{
 \includegraphics[scale = 0.5]{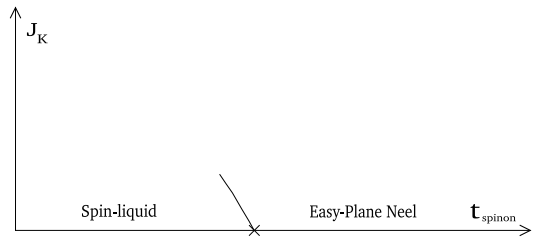}
}
\caption{Phase diagram in the presence of Kondo coupling. As shown in the text, weak Kondo coupling stabilizes the ordered Neel phase relative to the spin-liquid}
\label{phasedia}
\end{figure}

In the presence of weak coupling to the metallic electrons Ref. \cite{senthil03} argued that the quantum spin liquid state and its topological order are stable. The result is a `fractionalized Fermi liquid' phase which has a
$c$-electron Fermi surface coexisting with gapped fractionalized spinon excitations.
The transition to the Neel state has not been previously described and will be studied in this paper. We will focus on the interesting situation where the magnetic ordering wavevector can connect two points on the Fermi surface. We first argue that a weak coupling to the $c$-electrons actually stabilizes the Neel state relative to the
spin liquid state. Thus the phase boundary curves as shown in Fig. \ref{phasedia}.

 In the ordered state
the $c$-electrons near the points connected by the ordering wavevector will be strongly affected and the Fermi surface will change shape compared to the paramagnet. However we will argue that right at the magnetic quantum critical point the damping of the spin fluctuations due to the metallic quasiparticles is {\em irrelevant}. Thus the $c$-electrons dynamically decouple from the
critical spin fluctuations even though they are strongly coupled in the ordered state. This therefore provides an example of a strongly coupled magnetic quantum critical point in a metallic environment. We study the effect of the critical spin fluctuations on the $c$-electrons. Near `hot spots' on the Fermi surface (points that are connected to each other by the ordering wavevector) there is anomalous scattering leading to electron lifetimes that vanish as $T^{\overline\eta}$ with $\overline\eta > 1$. Thus the quasiparticle picture of the $c$-excitations is preserved at this quantum critical point. We describe the corrections to Fermi liquid results in transport and other quantities that result from the
anomalous scattering at these hot spots.

\section{Neel order - quantum spin liquid phase transition in Easy plane antiferromagnets}

Consider easy plane antiferromagnets with spin-$1/2$ per unit cell with collinear order at a wavevector $\vec Q$ on square lattice. The spin ordering pattern is characterized in terms of an $XY$ order parameter $\psi$
\begin{equation}
S^-(\vecb{r}_i) \sim e^{i\vecb{Q} \cdot \vecb{r}_i} \psi
\end{equation}
The $XY$ ordering allows for vortex topological defects characterized by an integer winding number that are point-like in two space dimensions.
Now consider disordering the $XY$ order to move into a quantum paramagnet by proliferating these vortices. When single vortices ({\em i.e} those with $2\pi$ winding) proliferate, the result is a quantum paramagnet with broken translational symmetry (a valence bond solid). Quantum spin liquid states result if the $XY$ order is killed by proliferating {\em paired} vortices ({\em i.e} those with $4\pi$ winding). The resulting spin liquid state is described as a gapped topologically ordered $Z_2$ spin liquid with bosonic spin-$1/2$ spinons, and a gapped $Z_2$ vortex (vison) that may be understood as an unpaired vortex. A lattice model exhibiting this physics was constructed in reference \cite{senthil02}.

As shown in reference \cite{senthil02}, the quantum phase transition between this spin liquid and the ordered state may be understood as a condensation of the bosonic spinons. In terms of the spinon fields (denoted as $b^\dagger$) it is in the universality class of the $2+1$ dimensional classical $XY$ model. However the physical spin order parameter is obtained as a square of the spinon field:
\begin{equation}
\psi = b^2
\end{equation}
Consequently the order parameter correlations decay with a large anomalous exponent:
\begin{equation}
<\psi^*(\bm{k},\omega) \psi(\bm{k},\omega)> \sim \frac{1}{\left( k^2 - (\omega + i \epsilon)^2\right)^{1-\overline\eta/2 }} \label{eq:suscep}
\end{equation}
with $\bar{\eta} = 1.45$ (see for example references \cite{biscal}). We would refer to this phase transition as lying in the $XY^{*}$ universality class in this paper since it is distinct from the usual $XY$ phase transition where the physical (planar) spin could be identified with the order parameter.

\section{Coupling to metallic electrons}
\subsection{Weak coupling phase diagram}

\begin{figure}[tb]
\centerline{
 \includegraphics[scale = 1.0]{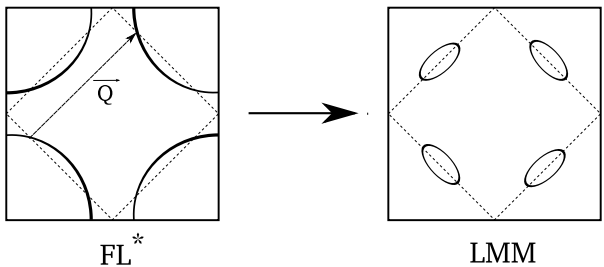}
}
\caption{Fermi Surface in the two phases $FL^{*}$ and $LMM$ (Local Moment Metal)}
\label{fig:fs}
\end{figure}

Now consider turning on a weak but non-zero interaction  of the form Eqn. \ref{kndo} between the insulating easy-plane magnet described above and a sea of non-interacting conduction electrons. We will focus on the effect of this interaction on the slow hydrodynamic variables of the insulating quantum magnet. These are simply the antiferromagnetic order parameter
$\psi$, the conserved uniform spin density $\vec L$, and the conserved energy density $\epsilon$. The order parameter coupling is the most important and takes the form
\begin{equation}
S_1 = \int_{\vecb{q}, \vecb{k_1}, \vecb{k_2}, \tau} J_K {\psi}^+(\vecb{q}). c^{\dagger}_{\vecb{k}_1}\vecb{\sigma^{-}}c_{\vecb{k}_2}\delta(\vecb{q} + \vecb{k}_1 - \vecb{k}_2 - \vecb{Q}) + c.c.
\end{equation}
where $\vecb{Q}$ is the ordering wavevector.
We will assume that $\vecb{Q}$ can connect two distinct points on the conduction electron Fermi surface \ref{fig:fs}. The electrons near such Fermi surface points scatter strongly of the fluctuations of the magnetic order parameter. In the ordered phase this will open up a gap at such points of the conduction electron Fermi surface. The size of the gap $\Delta_{el}$ will be determined by the strength of order parameter $\psi_0 = |\psi|$.
Within mean field theory $\Delta_{el} \sim J_K \psi_0$. In the terminology of Ref. \cite{svs} this is a `local moment magnetic metal' (LMM).

In the disordered spin liquid phase the order parameter fluctuations are gapped. Consequently they can be integrated out in favor of a short-ranged spin fluctuation mediated electron-electron interaction. In the absence of any special nesting condition on the Fermi surface (which we assume) this does not significantly affect the
low energy properties of the $c$-electrons. Further the volume of the Fermi surface is determined by the density of $c$-electrons alone without counting the local moments. As argued in Ref. \onlinecite{senthil03}, the fractionalization and topological order of the local moment system survives the coupling to the metal. This metallic phase is a fractionalized Fermi liquid ($FL^*$) which has spinon and associated topological excitations coexisting with a small Fermi surface of $c$-electrons.

One important issue is the relative stability of the spin liquid phase compared to the Neel phase in the metallic environment. In other words does the coupling to the conduction electrons enhance or decrease the parameter regime in which the spin liquid exists?  We argue now that at least at weak coupling the Neel state gains in stability relative to the spin liquid. Physically this may be traced to the damping of the spin fluctuations
due to particle-hole pairs of the Fermi liquid. In the paramagnetic phase, if we integrate out the conduction electrons, the (imaginary time) action for the Neel order parameter acquires the familiar Landau damping term $\int_{\vec q, \omega} |\omega| |\psi|^2$. To assess the effect of this damping on the phase diagram, consider the following simple model for the spinons $b$ on a $2+1$ dimensional space-time cubic lattice with the action
\begin{eqnarray}
S & = & S_0 + S_1 \nonumber \\
& = & \sum_{<(\bm{x},\tau),(\bm{x'},\tau')>} -t\left[b^{\dagger}(\bm{x},\tau) b(\bm{x}',\tau') + h.c. \right] \nonumber \\
& & + g \sum_{\bm{x},\tau,\tau'} K(\tau-\tau') b^2(\bm{x},\tau). b^{\dagger 2}(\bm{x},\tau') \label{action_phase}
\end{eqnarray}
subject to the constraint $b^{\dagger}(\bm{x},\tau)b(\bm{x},\tau) = 1$.   Here $K(\tau) = \frac{1}{{\tau}^2}$ is the imaginary time Fourier transform $|\omega|$. The constant $g \approx {J_K}^2/{v_F Q}$ where $v_F$ is the Fermi velocity. Note that the coupling term breaks the rotational invariance in $\bm{x}-\tau$ space. As $t$ increases, at certain critical value the $b$ field orders. Denoting this critical value in the absence and presence of $g$ by $t_{0c}$ and $t_{c}$ respectively, we find that $t_{c} = t_{0c} - \alpha g$ where $\alpha = O(1)$ is a positive constant. The details are given in the appendix A.

\subsection{Critical properties}
We now turn to  the critical properties of the transition in the presence of a weak coupling to the conduction electrons.
The electrons will couple to the slowly varying component of the order parameter field. The lowest order addition to the Hamiltonian  consistent with the symmetries of the lattice and spin-rotation is
\begin{equation}
\Delta{H_1} = J_K\int_{\bm{r}}c^{\dagger}(\bm{r}){\sigma^{+}}c(\bm{r}).\bm{S^{-}(\bm{r})} + h.c.
\end{equation}
where `$g$' is a coupling constant.  Another potentially important term involves coupling between energy  densities of the electron and local moments. To leading order this takes the form
\begin{equation}
\Delta{H_2} = \lambda \int_{\bm{r}}\left[ c^{\dagger}(\bm{r})c(\bm{r})\right] .\left[ b^{\dagger}(\bm{r})b(\bm{r})\right]
\end{equation}
with $\lambda$ as another coupling constant. Let us now integrate out the conduction electrons to obtain an effective action for the spinon fields.
As already described above, the main effect of $\Delta{H_1}$ is to generate the usual Landau damping term in the action $S_1 = \int_{\kw} g|\omega||\psi(\bm{k},\omega)|^2$. From the point of view of renormalization group (RG), one crucial observation concerning $S_1$ is that the kernel $K(\omega)$ is long-ranged in the time-domain ($K(\tau) = \frac{1}{\tau^2}$) and thus couples $\psi$'s at two separate points in space-time. This implies that at the tree-level, one could directly use the scaling dimensions for individual $\psi$ operator's even though $|\psi(\kw)|^2$ is a composite operator. Using the RG transformation appropriate for the critical fixed point in the absence of the conduction electrons, $\omega \rightarrow s \omega, \bm{k} \rightarrow s \bm{k}, \psi \rightarrow s^{(\overline{\eta}-5)/2}$ with $s < 1$ implies

\be
S_1 \rightarrow s^{\overline{\eta}-1} \int_{\kw} g |\omega||\psi(\bm{k},\omega)|^2
\ee

Thus the damping coefficient $g$ flows  as

\begin{equation}
\frac{dg}{dl} = (1-\overline{\eta})\,g
\end{equation}

Since $\overline{\eta}>1$, this term is clearly irrelevant for $g$ in the vicinity of the unperturbed fixed point
Next consider $\Delta{H_2}$. Proceeding in the exactly similar manner, one integrates out the electrons and obtain an effective term $g' \int_{\kw} \frac{|\omega|}{k} |b(\bm{k},\omega)|^2|b(\bm{-k},-\omega)|^2$ where $g' \approx \lambda^2/E_F$. A simple scaling analysis \cite{sachdev02} similar to above yields
\begin{equation}
\frac{dg'}{dl} = (-3+2/\nu)\,g'
\end{equation}

where $\nu$ is the critical exponent associated with scaling of specific heat. Since $\nu > 2/3$ at $O(2)$ critical point \cite{mass01}, $\Delta{H_2}$ also turns out to be irrelevant.

Thus we have the remarkable result that a weak coupling to the conduction electrons does not change the critical properties of the transition from that in the insulating magnet. This is in striking contrast to the HMM theory where the Landau damping is strongly relevant. Thus in the present theory the spin fluctuations are described by a non-trivial interacting fixed point, and hence satisfy scaling (for instance in $\omega/T$). In contrast due to the Landau damping, the HMM theory is right at the upper critical dimension in $d = 2$ (and the fixed point is Gaussian).


\section{Effect of Coupling on Conduction Electrons}
Though the coupling to AFM order parameter turns out to be irrelevant for the phase transition critical properties, it may still affect the electronic properties significantly.
First consider the conduction electrons' band structure in the magnetically ordered side. At the mean field level, the effect of coupling could be captured by the following Hamiltonian

\begin{equation}
H = \epsilon_k c^{\dagger}_{\bm{k}} c_{\bm{k}} + J_K \, N_0 \left( \, c^{\dagger}_{\vecb{k+Q}}\vecb{\sigma_x}c_{\vecb{k}} + h.c.\right)
\end{equation}

Here $N_0$ is the AFM order parameter and we have assumed that spins order along $\bm{\hat{x}}$ direction. The above Hamiltonian is easily diagonalized and one obtains two bands $E^{\pm} = \frac{\epsilon_{\bm{k}} + \epsilon_{\bm{k+Q}}}{2} \pm \sqrt{\left( \frac{ \epsilon_{\bm{k}} - \epsilon_{\bm{k+Q}} }{2}\right)^2 + J_K^2 N_0^2} $. For any non-zero $N_0$, the unit cell is doubled and thus one finds electron and hole pockets emerging on the ordered side \cite{sachdev95, chub97}. As $N_0$ increases, the hole pockets grow in size at the expense of electron pockets, the difference in areas being constant and determined by the density of conduction electrons \cite{sachdev95, chub97}. For small $N_0$, the separation in momentum space between hole and electron pockets grows in proportion to $N_0 = \left\langle \ b^2 \right\rangle \sim |t-t_c|^{\overline{\beta}}$ where $t$ is the tuning parameter for transition and $\overline{\beta} \approx 0.83 $ \cite{biscal}. Thus the separation grows much slower compared to a conventional HMM scenario where the AFM order
$N_0 \sim |t-t_c|^{\beta}$ with $\beta \sim 0.35$ corresponding to 3-D X-Y universality class.

Though potentially there are many other interesting questions to be answered on the magnetically ordered side (e.g. the limit $N_0 \rightarrow 0$ and $k,\omega \rightarrow 0$ for various correlation functions may not commute \cite{sachdev95}), our focus in the remaining paper would be on the quantum critical regime. We begin by calculating single-particle Green's function for conduction electrons in the quantum critical regime.

\subsection{Single-Particle Green's Function}
Qualitatively, since critical fluctuations are centered around the ordering wavevector $\bm{Q}$, the scattering rate at the Fermi points connected by $\bm{Q}$ will get enhanced and one would expect that the electron transport properties might show signature of such `hot spots'.
One simple and useful calculation one might do is to calculate the one-particle Green's function $G(\bm{k},i\omega_n)$
for a conduction electron perturbatively in the coupling $g$.

\begin{figure}[tb]
\centerline{
 \includegraphics[scale = 0.5]{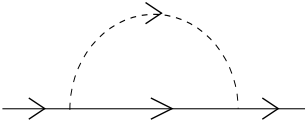}
}
\caption{Lowest order contribution to the electron's self-energy. The full line is the electron's non-interacting Green's function while the dashed line denotes spin-spin correlator $\chi(\bm{k},\omega)$}
\label{fig:selfe}
\end{figure}

The lowest order term that contributes to the self energy $\Sigma(\bm{k},i\omega_n) = {G_0^{-1}(\bm{k},i\omega_n)}-
{G^{-1}(\bm{k},i\omega_n)}$ (see Fig. \ref{fig:selfe}) is

\begin{equation}
\Sigma(\bm{k},i\omega_n) = T J_K^2  \sum_{\bm{k'},i\Omega_m} \chi_s(\bm{k'},i\Omega_m) G_0(\bm{k}-\bm{k'},i\omega_n-i\Omega_m) \label{eq:selfe1}
\end{equation}

where $\chi_s(\bm{k},\omega) = \left\langle S^{+}(\bm{k}, \omega) S^{-}(\bm{k}, \omega)\right\rangle = \left\langle \psi^{*}(\bm{k-Q}, \omega) \psi(\bm{k-Q}, \omega)\right\rangle$ is the dynamic susceptibility of local moments and can be read off from Eqn. ~\ref{eq:suscep} while $G_0(\bm{k},i\omega_n) = \frac{1}{i \omega_n - \epsilon_{\bm{k}}}$ is electron's unperturbed Green's function. Note that here $i\omega_n$ is a fermionic while $i\Omega_m$ is a bosonic Matsubara frequency.

It is useful to go to the  spectral representation for $\chi(\bm{k},\Omega)$

\begin{eqnarray}
\chi(\bm{k},i\Omega_m) & = & \int \frac{A_b(\bm{k},\omega) d\omega}{\omega - i\Omega_m} \,\,\,\,\,\, \mathrm{where} \nonumber \\
A_b(\bm{k},\omega) & = & C \frac{\theta(\omega^2-c^2{|\bm{k}-\bm{Q}|}^2)}{{(\omega^2-c^2{|\bm{k}-\bm{Q}|}^2)}^{1-\overline{\eta}/2}} \label{eq:A_b}
\end{eqnarray}

where $C$ is a constant of $O(1)$. Substituting ~\ref{eq:A_b} and expression for $G_0(\bm{k},i\omega_n)$ in ~\ref{eq:selfe1}, one obtains

\begin{equation}
\Sigma(\bm{k},i\omega_n) \approx J_K^2 \int_{\omega,\bf{q}}\frac{\left[ n_b(\omega^{'})+1-n_f(\epsilon_{\bm{k}+\bm{q}+\bm{Q}})\right]\theta({{\omega}^{'}}^2-q^2) }{\left[(i\omega_n-\omega^{'}-\epsilon_{\bm{k}+\bm{q}+\bm{Q}})(\omega^{'2}-c^2 q^2)^{1-\overline\eta/2}\right]} \label{eq:selfe2}
\end{equation}

where $n_b$ and $n_f$ are Bose and Fermi functions respectively and $\epsilon_{\bm{k}} \equiv E_{\bm{k}} - \mu$ is the single-particle energy. On physical grounds it is expected that $\textrm{Im} \,\Sigma(\bm{k},i\omega_n)$  would be significant only for values of $\bm{k}$ such that both $\bm{k}$ and $\bm{k + Q}$ lie on the Fermi surface. Such points on the Fermi surface are generally called `hot spots'. Let $\bm{k_0}$ be one such point. We find (Appendix B) that at $T = 0$,

\bea
  \mathrm{Im}\,\Sigma(\bm{k},i\omega_n= \omega ^{+} ) \sim  J_K^2 \,sign(\omega)\, \omega^{\overline\eta}
\eea

for $|\bm{k}-\bm{k_0}| \lesssim \omega/c $ \newline

while

\be
 \mathrm{Im}\,\Sigma(\bm{k},i\omega_n=\epsilon_{\bm{k}}^{+} = 0) \sim  J_K^2 \, T^{\overline\eta}
\ee

at a non-zero temperature $T$ for $\bm{k} \in$ the Fermi Surface and satisfying $|\bm{k}-\bm{k_0}| \lesssim T/c $. For points $\bm{k}$ near the Fermi surface away from the hot spots the self-energy has the usual Landau Fermi liquid form, namely, $\textrm{Im}\, \Sigma(\bm{k},i\omega_n=\omega^{+}) \sim \omega^2 + T^2$. Evidently, the hot-spots acquire a width $O(T)$ at non-zero but small temperature $T$.
Perhaps most importantly, the electronic quasiparticle maintains its integrity at the transition at all points on the Fermi surface since ${\frac{\partial{\textrm{Im} \,\Sigma(\omega)}}{\partial\omega}}|_{\mathrm{hot \, \, spot}} \sim {\omega}^{\overline\eta-1} \rightarrow 0$ as $\omega \rightarrow 0$.

Finally, the real part of self-energy at $T=0$ for points near a hot-spot behaves as $\mathrm{Re}\, \Sigma(\bm{k} \approx \bm{k_0},\omega) \sim J^2_K\left(  -\omega \Lambda^{\overline{\eta}-1} + \omega^{\overline\eta} \right) $ where $\Lambda \sim v_F/a $ is an ultraviolet frequency cut-off and $a$ is the lattice spacing (Appendix B).

\subsection{Thermodynamic and Transport properties of Electronic Quasiparticles}

\noindent \underline{ {\it Specific heat}}:
The correction to the internal energy of the electronic system is given by

\be
\Delta U = \sum_{\bm{k}} \Delta \epsilon_{\bm{k}} \, n_f(\epsilon_{\bm{k}})
\ee

where $\Delta \epsilon_{\bm{k}} = \textrm{Re}\, \Sigma(\bm{k}, i\omega_n=\epsilon_{\bm{k}}^{+})$ is the correction to the single-particle energy levels. As argued above that the regions which contribute primarily to $\Delta \epsilon_{\bm{k}}$ are located around hot-spots (denoted $\bm{k_0}$ here and above) and have a width proportion to temperature $T$. Setting up a polar coordinate system $q-\phi$ near one such $\bm{k_0}$ and noticing that $\epsilon_{\bm{k}} \approx v_F q \,\textrm{cos}(\phi)$ , the expression for $\Delta U $ may be written as

\be
\Delta U \approx \int_0^{2\pi} d\, \phi \int_0^T q \,dq \frac{J^2_K \left[\Lambda^{\overline{\eta}-1} q cos(\phi) - \left( q cos(\phi)\right)^{\overline\eta}  \right] }{e^{\beta q cos(\phi)} + 1}
\ee

which immediately yields $\Delta U \approx J^2_K \left( \Lambda^{\overline{\eta}-1} T^3 - T^{\overline\eta + 2} \right) $ just from scaling. Thus the correction to electronic specific heat $\Delta C_v$ is

\bea
\Delta C_v & = &\frac{\partial U}{\partial T} \\
& \approx & J^2_K \left( \Lambda^{\overline{\eta}-1} T^2 - T^{\overline\eta + 1} \right)
\eea

which has the same temperature dependence as the contribution from the local moments ($\propto T^2$) at low temperatures while being subdominant to the contribution from the electronic system in the  absence of any coupling to spin fluctuations ($\propto T$) at low temperatures.

\noindent \underline{ {\it Spin Susceptibility}}:


The dominant contribution to the spin-susceptibility comes from the critical spin fluctuations $\chi_s$. Here we calculate the subleading contribution $\chi_e$ arising from the conduction electrons. The spin-susceptibility for a non-interacting Fermi system $\chi_{eo}(\bm{k}, \omega)$ at $T = 0$ for $\omega \ll k$ is given by $\chi_{eo}(\bm{k},\omega) \approx -N(E_F)(1 + i C \omega)$ where $C$ is constant. To $O(J_K^2)$, the correction to this result due to coupling to critical fluctuations could be calculated by replacing the non-interacting Green's function by the full interacting one in the `polarization bubble' diagram for the free electron susceptibility.

Thus
\begin{equation}
\chi_e(\bm{k},i\omega_n) = T \sum_{\bm{q},i\nu_r}G(\bm{k}+\bm{q},i\omega_n+i\nu_r)G(\bm{q},i\nu_r)
\end{equation}

Expressing the Green's function in terms of spectral function $A(\bm{k},\omega)$ and doing the summation over Matsubara frequencies $i\nu_r$

\begin{equation}
\chi_e(\bm{k},i\omega_n) = \int_{\bm{q}}\int_{\Omega_1,\Omega_2} \left[ \frac{n_f(\Omega_1)- n_f(\Omega_2)}{i\omega_n+\Omega_2-\Omega_1}\right] A(\bm{k}+\bm{q},\Omega_1)A(\bm{q},\Omega_2) \label{eq:chi_full}
\end{equation}

The spectral function $A(\bm{k},\omega) = \textrm{Im} G(k, \omega^{+})$ is expressed simply in terms of inverse single-particle life-time $\Gamma(\bm{k},\omega)$ as

\begin{equation}
A(\bm{k},\omega) = \frac{\Gamma(\bm{k},\omega)}{{\Gamma}^2(\bm{k},\omega) + {(\omega - \epsilon'_{\bm{k}})}^2}
\end{equation}

where $\epsilon'_{\bm{k}} = \epsilon_{\bm{k}} + \textrm{Re}\, \Sigma(\bm{k},\omega = \epsilon_{\bm{k}}^{+}) \propto \epsilon_{\bm{k}} $ to the leading order.
The imaginary part of $\chi(\bm{k},\omega)$ is thus obtained as

\begin{equation}
\textrm{Im} \chi_e(\bm{k},\omega) = \int_{\bm{q}}\int_{\Omega}\left[n_f(\Omega)-n_f(\Omega-\omega) \right] A(\bm{k}+\bm{q},\Omega) A(\bm{q},\Omega-\omega)
\end{equation}

Since $A(k,\omega)$ is modified significantly due to critical fluctuations only near hot points, the above integral would get important contribution beyond the free electron result only if $\bm{k} \approx \bm{Q}$. Further, the momentum integration over $\bm{q}$ would get significant contribution only from a thin region around hot-spots of linear dimensions $\omega$. Using the fact that for $\epsilon_{\bm{k}} \lesssim \omega$ and $\bm{k}$ belonging to hot-spot, $A(\bm{k}, \omega) \sim 1/\omega^{2-\overline\eta}$, one readily obtains $\textrm{Im} \, \chi_e(k\approx Q, \omega) \sim \textrm{Im} \, \chi_{eo}(k\approx Q, \omega) + \omega^{2 \overline\eta -1}$. Since $\overline\eta > 1$, this implies that to the leading order $\chi_e = \chi_{eo}$ and thus the leading correction to the total spin-susceptibility is dominated by the free-electron susceptibility.

\noindent \underline{ {\it DC Conductivity}}:
Under the relaxation time approximation for the linearized Boltzmann's transport equation in the presence of an electric field $\bm{E}$, the conductivity for a Fermi liquid at low temperatures is given by $\sigma = \frac{ne^2 \left\langle \tau_{\bm{k}}\right\rangle_{FS} }{m}$ where $n$ is the carrier density and $\left\langle \tau_{\bm{k}}\right\rangle_{FS} $ is the relaxation time $\tau_{\bm{k}}$ averaged over the Fermi Surface. The relaxation time $\tau$ is proportional to the inverse scattering rate but in general, one needs to weight the scattering rate by the factor $(1-cos(\theta))$ where $\theta$ is the scattering angle. Here we neglect this factor since the scattering due to spin excitations have $\theta \sim \theta_0 = O(1)$ where $\theta_0$ is the angle subtended by the hot spots $\bm{k_0}$ and $\bm{k_0 + Q}$ at the origin of the Brillouin zone. Further the leading contributions of the other two scattering processes we consider below namely electron-electron and impurity scattering is unchanged by the introduction of this factor.

Near the hot points the scattering rate $1/\tau$ of electronic quasiparticles due to critical spin fluctuations is proportional to $ \textrm{Im}\, \Sigma \sim T^{\overline\eta}$. Thus $\tau_M/\tau \approx x + (T/\Gamma)^{\overline{\eta}} \equiv x + t^{\overline{\eta}}$ where $x$ measures the scattering rate due to (weak) disorder and is isotropic while $\tau_M$ and $\Gamma$ are characteristic scattering time and energy scale respectively. In the cold regions one expects typical Fermi liquid behavior, thus $\tau_M/\tau \approx x + \left( T/\Gamma\right) ^2 = x + t^2$. From this and using the fact that the width of the hot spots is proportional to $t$, one obtains the following expression for the conductivity within the linearized relaxation time approximation

\be
\sigma \propto \frac{t}{x + t^{\overline{\eta}}} + \frac{1-t}{x + t^2}
\ee

Following Rosch \cite{rosch99}, we consider the two limits, $x < t^2 <1$ and $ t < x < 1$. We find that in both limits the resistivity $\rho = 1/\sigma \propto x + t^2$. Thus the Fermi liquid result $\rho \propto T^2$ remains valid at low temperatures. This result is very different compared to that for the resistivity near the $z = 2$ quantum critical point described within Hertz-Millis-Moriya theory  where one obtains $\rho \propto x + T^2$ in the limit $ x < t^2 < 1 $ while it is \textit{anomalous} and $\propto x + T^{3/2}$ in the limit $t < x < 1$ \cite{rosch99}.

\section{Summary and Discussion}
A crucial ingredient in the example presented in this paper for the magnetic transition in the presence of metallic coupling has been the fact that 1)  For the parent insulating system, the spinons are deconfined at the phase transition and that 2) the critical exponent $\overline\eta >1$. We found that due to these facts conduction electrons become dynamically decoupled from the underlying spin-system at low energies and therefore the phase transition remains in the exotic universality class $XY^{*}$ even in the presence of coupling to conduction electrons. Further we found that the metallic coupling favors the ordered phase over paramagnetic spin-liquid. We also analyzed the effect of critical fluctuations on the conduction electrons and found that the Landau quasiparticle survives at the phase transition. This enabled us to calculate corrections to various thermodynamic and transport properties of the conduction electrons due to critical fluctuations. Overall, we found that to the leading order the critical fluctuations do not modify any thermodynamic or transport property of the conduction electrons and all the results from standard Fermi liquid theory remain valid.

Contrasting our example with the conventional HMM \cite{hertz76} theory, one immediately notices rather stark differences. The HMM theory describes the onset of spin density wave order  from a paramagnet Heavy Fermi Liquid (HFL) phase with a large Fermi surface. Spinon excitations do not exist in either phase. The Landau damping of the spin density wave fluctuations leads to a dynamic critical exponent $z = 2$. The HMM theory is then right at the upper critical dimension in $d = 2$, and the physics is determined by marginal terms associated with interactions between the order parameter fluctuations. Non-trivial scaling of the order parameter fluctuations is not expected within the HMM theory. On the other hand, in our example, the paramagnetic  metallic phase has a small Fermi surface and gapped deconfined spinon excitations though the magnetically ordered phase is a conventional phase with no spinon excitations. Nevertheless the critical theory for the transition is most usefully formulated in terms of the spinon degrees of freedom. The  theory has $z = 1$ and is below  its upper critical dimension. Hence the order parameter fluctuations naturally   naturally exhibit $\omega/T$ scaling for various critical properties.

In summary, we provide an example of a magnetic phase transition in a metallic system where the conventional HMM theory fails and which at the same time shares some of the properties with rather intriguing phases \cite{fried09} and phase transitions \cite{stock98, schr98, schr00} in itinerant metallic systems.

\textit{Acknowledgement}: TS was supported by NSF Grant DMR-0705255.

\section{Appendix A}

Here we analyze the phase diagram for action $S$ is eqn. \ref{action_phase} for $g \ll 1$. Rewriting it as

\bea
& & S = \sum_{<(\bm{x},\tau),(\bm{x'},\tau')>} -t\left[b^{\dagger}(\bm{x},\tau) b(\bm{x}',\tau') + h.c. \right] \nonumber \\
& & + g \sum_{\bm{x},\tau,\tau'} K(\tau-\tau') \left[ b(\bm{x},\tau). b^{\dagger}(\bm{x},\tau')\right] \left[  b(\bm{x},\tau). b^{\dagger}(\bm{x},\tau')\right] \nonumber
\eea

The above form is amenable to a large-$N$ generalization which allows us to do a more controlled calculation. Thus we consider $N$ species of bosons $b_{\alpha}$ with the following action $S_N$:

\bea
& & S_N = \sum_{<(\bm{x},\tau),(\bm{x'},\tau') > \alpha} -t\left[b^{\dagger}_{\alpha}(\bm{x},\tau) b_{\alpha}(\bm{x}',\tau') + h.c. \right] \nonumber \\
& & + \frac{g}{N} \sum_{\bm{x},\tau,\tau',\alpha, \beta} K(\tau-\tau') \left[b_{\alpha}(\bm{x},\tau). b_{\alpha}^{\dagger}(\bm{x},\tau')\right]^2 \nonumber
\eea

Thus the $O(2)$ symmetry of the original action has been replaced with $O(2N)$ symmetry. At $N= \infty$, the above action reduces to the simple form,

\bea
& & S_N  = \nonumber \\
& &\sum_{\bm{k},\omega,\bm{k'},\omega', \alpha} b^{\dagger}_{\alpha}(\bm{k},\omega) b_{\alpha}(\bm{k},\omega) \left[  \epsilon(\bm{k},\omega)
 + g K(\omega + \omega') G_{\alpha}(\bm{k'},{\omega}') \right] \nonumber
\eea

Here $K(\omega) = |\omega|$ is the Fourier transform of $K(\tau)$, $\epsilon(\bm{k},\omega) = -2t\left( cos(k_x) + cos(k_y) + cos(\omega) \right)$  and $G_{\alpha}(\bm{k},\omega) = \left\langle b_{\alpha}(\bm{k},\omega)^{\dagger} b_{\alpha}(\bm{k},\omega) \right\rangle$. Since we are approaching the phase transition from the paramagnetic side, $G_{\alpha}(\bm{k},\omega) \equiv G(\bm{k},\omega)$ is independent of $\alpha$.

First consider $S_N$ at $g = 0$. Imposing the constraint $b^{\dagger} b = 1$ using a Lagrange multiplier $\lambda$, the action becomes,

\be
S_0 = \sum_{\bm{k},\omega,\alpha} b^{\dagger}_{\alpha}(\bm{k},\omega) b_{\alpha}(\bm{k},\omega) \left[  \epsilon(\bm{k},\omega) + \lambda \right]
\ee

Imposing self-consistency for $G(\kw)$ yields

\be
 \sum_{\kw} \frac{1}{\lambda -2t\left( cos(k_x) + cos(k_y) + cos(\omega) \right)} = 1
\ee

At the phase transition, the gap to excitations closes and hence $\lambda = 6t$. Thus the critical $t = t_{0c}$ at $ g = 0$ is given by

\be
t_{0c} = \sum_{\kw} \frac{1}{6 -2\left( cos(k_x) + cos(k_y) + cos(\omega) \right)}
\ee

Next consider the self-consistency condition at the phase transition for a non-zero but small $g$. One finds

\be
 \sum_{\kw} \frac{1}{6t_c -2t_c\left( cos(k_x) + cos(k_y) + cos(\omega) \right) + g F(\omega)} = 1
\ee

where $F(\omega) = \sum_{\bm{k'},\omega'}\left\lbrace \left[ K(\omega + \omega') - K(\omega')\right]  G(\bm{k'},{\omega}')\right\rbrace$ Using the expression for $G$ at $g = 0$, one readily finds that $F(\omega)  > 0 \,\,\forall \,\,\omega$ . Thus to $O(g)$, one reaches the conclusion that $t_c = t_{0c} - \alpha g$ where $\alpha$ is a positive constant. Thus the phase boundary curves as shown in Fig. \ref{phasedia}.

\section{Appendix B}

Using eqn. \ref{eq:selfe2} ,  The imaginary part of the self-energy is

\bea
& & \mathrm{Im} \, \Sigma(\bm{k},i\omega_n = \omega^{+}) \approx  \nonumber \\
& &  J_K^2 \int_{\omega',\bf{q}}\frac{\left[ n_b(\omega^{'})+1-n_f(\epsilon_{\bm{k}+\bm{q}+\bm{Q}})\right]\theta({{\omega}^{'}}^2-c^2q^2)  }{(\omega^{'2}-c^2 q^2)^{1-\overline\eta/2}} \times \nonumber \\
& & \delta(\omega - \omega' - \epsilon_{\bm{k + Q + q}}) \label{eq:imgselfe}
\eea

First consider $\mathrm{Im} \Sigma(\bm{k_0},i\omega_n = \omega^{+})$ at $T = 0$ where $\bm{k_0}$ is a hot spot i.e. both $\bm{k_0}$ and $\bm{k_0 + Q}$ lie on the Fermi surface. Let's resolve $\bm{q}$ in components $q_\parallel$, $q_{\perp}$ parallel and perpendicular to the Fermi surface respectively. Since the important contribution at small $\omega$ comes from $\bm{q}$ located near the Fermi surface, $\epsilon(\bm{k_0 + Q+ q}) \approx v_F q_\perp$. Substituting this and the expressions for $n_b$ and $n_f$ in $T \rightarrow 0$ limit into the equation \ref{eq:imgselfe} one finds,

\bea
& & \mathrm{Im} \, \Sigma(\bm{k_0},i\omega_n = \omega^{+}) \approx J_K^2  sign(\omega) \times \nonumber \\
& &  \int dq_\parallel dq_\perp \frac{\theta(\omega - v_F q_\perp) \theta(v_F q_\perp)\theta\left( (\omega - v_F q_\perp)^2-c^2q^2\right)}{\left( (\omega - v_F q_\perp)^2-c^2q^2\right) ^{1-\overline\eta/2}} \nonumber \\
\eea

The integration over $q_\parallel$ could be done easily by just rescaling the variables yielding

\bea
& & \mathrm{Im} \, \Sigma(\bm{k_0},i\omega_n = \omega^{+}) \nonumber \\
& &  \approx J_K^2 \, sign(\omega)\,\int_0^{\frac{\omega}{v_F+c}} dq_\perp {(({\omega}-v_F q_\perp) }^2-c^2{q_\perp}^2)^{(\overline\eta-1)/2} \nonumber \\
& & \approx J_K^2 \,sign(\omega)\, \omega^{\overline\eta} \nonumber
\eea

Similarly consider $\mathrm{Im} \Sigma(\bm{k},i\omega_n = \omega^{+})$ when $\bm{k}$ is located away from hot-spots which implies $\epsilon_{\bm{k+Q}} {\not=} 0$. Following the same procedure as above, one finds that the Heaviside function imposes the condition that for $\mathrm{Im} \, \Sigma$ to be non-zero, $\omega \gtrsim \epsilon_{\bm{k+Q}}$. For $\bm{k}$'s that do satisfy this condition, $\textrm{Im} \Sigma$ could be approximated by the same expression as above and is thus $\propto \overline\omega^{\eta}$ for small $\omega$.

Next we consider $\mathrm{Im} \Sigma(\bm{k_0},i\omega_n = \epsilon_{\bm{k_0}}^{+} = 0)$ at finite but small temperatures.

\bea
& & \mathrm{Im} \, \Sigma(\bm{k_0},i\omega_n = 0^{+}) \approx \nonumber \\
& & J_K^2 \int_{\bf{q}} \frac{ \left[ n_b(-\epsilon_{\bm{k_0}+\bm{q}+\bm{Q}}) + 1 - n_f(\epsilon_{\bm{k_0}+\bm{q}+\bm{Q}})\right] \theta ({\epsilon^2_{\bm{k_0}+\bm{q}+\bm{Q}}-c^2q^2) }} {( \epsilon^2_{\bm{k_0}+\bm{q}+\bm{Q}}-{c^2q^2}) ^{1-\overline\eta/2}} \nonumber
\eea

The naive approximation $\epsilon(\bm{k_0 + Q+ q}) \approx v_F q_\perp$ yields zero for the above integral because such an approximation leads to a spurious symmetry $q_\perp \leftrightarrow -q_\perp$. It is easy to verify that using the full expression for $\epsilon(\bm{k_0 + Q+ q}) = q^2/2m + v_F q_\perp$ the integrand is non-zero as it should be on physical grounds. The scale dependence on $T$ could be easily extracted by rescaling the variables and one finds,
\bea
\mathrm{Im} \, \Sigma(\bm{k_0},i\omega_n = 0^{+}) & \sim & J_K^2 T^{\overline\eta}
\eea

At points that are located away from the hot-spots, Fermi/Bose functions in the above equation impose that $\mathrm{Im} \, \Sigma(\bm{k_0},i\omega_n = 0^{+}) \approx 0 $ if  $|\bm{k- k_0}| \gtrsim T/c$.

Finally the real part of self-energy for points in the vicinity of hot-spots at finite temperature could be calculated from the corresponding imaginary part using the Kramers-Kronig relation,

\bea
\mathrm{Re} \, \Sigma(\bm{k} \approx \bm{k_0},\omega) & = & \frac{1}{\pi} \textrm{P} \int_\nu \frac{|\textrm{Im}\, \Sigma(\bm{k_0},\nu )|}{\omega -\nu}
\eea

where the symbol $\textrm{P}$ denotes the Principal Value of the integral. Substituting $\textrm{Im} \, \Sigma(\bm{k_0},\nu) \approx J_K^2 sign(\nu) \, {\nu}^{\overline\eta}$ one obtains $\mathrm{Re}\, \Sigma(\bm{k} \approx \bm{k_0},\omega) \sim J^2_K\left(  -\omega \Lambda^{\overline{\eta}-1} + \omega^{\overline\eta} \right) $ where $\Lambda$ is an ultraviolet frequency cut-off.

\end{document}